\documentclass[12pt,preprint]{revtex4}
\usepackage{amsmath}
\usepackage{graphicx}
\usepackage{amssymb, graphics}
\usepackage{amsfonts}

\newcommand{\rd}{\mathrm{d}}

\begin{document}
\title{Bulk viscous cosmology: unified dark matter}
\author{Xu Dou}
\email{dowxdou@gmail.com}
\author{Xin-He Meng}
\email{xhm@nankai.edu.cn}

\affiliation{School of physics, Nankai University, Tianjin 300071,
China}

\begin{abstract}
The bulk viscosity is introduced to model unified dark matter. The
viscous unified model assumes the universe is filled with a single
fluid with the bulk viscosity. We review the general framework of
the viscous cosmology. The Hubble parameter has a direct connection
with the bulk viscosity coefficient. For concrete form of the bulk
viscosity, the Hubble parameter which has the scaling relation with
the redshift can be obtained. We discuss two viscosity models and
the cosmological evolution to which they lead. Using SNe Ia data,
the viscosity model can be fitted. We briefly review the fitting
method here.
\end{abstract}

\maketitle

\section{Introduction}

Both dark matter problem and the cosmic acceleration problem
challenge physicists' understanding of the universe. In the standard
$\Lambda$CDM model, two mixed fluids, dark matter and dark energy
fluid, are assumed. These two fluids influence the cosmic evolution
separately. However, present gravitational probe does not have the
ability to differentiate these two fluids. This is the dark
degeneracy problem \cite{dg1} \cite{dg2}. It is reasonable to model
dark matter and dark energy with single fluid or single field
assumption. Some unified models have been proposed to detect the
possibility of this unified assumption, like unified dark fluid
model \cite{sf1} \cite{sf2} \cite{sf3} \cite{sf4} \cite{sf5}, which
assumes the single fluid equation of state; Chaplygin gas model and
generalized Chaplygin gas \cite{cp} \cite{cp2} \cite{cp3} \cite{cp4}
\cite{cp5}, which discuss the cosmology consequences of an exotic
equation of state; scalar field method \cite{field1} \cite{field2}
\cite{field3}.

The introduction of viscosity into cosmology has been investigated
from different view points \cite{G} \cite{PC}. There are some recent
developments like dark energy model \cite{vd1} \cite{vd2} \cite{vd3}
\cite{vd4}, the cosmic singularity \cite{vs1}. In this review, we
give a brief introduction to unify dark matter and dark energy with
viscosity medium. In such models, the universe is assumed to be
filled with viscous single fluid \cite{sv1} \cite{sv2} \cite{sv3}
\cite{sv4} \cite{sv5} \cite{sv6} \cite{sv7}. The cosmic density is
not separated as dark energy part and dark matter part. The bulk
viscosity contributes to the cosmic pressure, and plays the role as
accelerating the universe. After considering the bulk viscosity, the
cosmic pressure can be written as
\begin{equation}
p=(\gamma-1)\rho-3\zeta H
\end{equation}
Where $\gamma$ parameterizes the equation of state. Generally the
form of bulk viscosity is chosen as a time-dependent function. In
\cite{v1} \cite{v2} \cite{v3} \cite{v4}, a density-dependent
viscosity $\zeta=\alpha\rho^{m}$ coefficient is investigated
extensively. For modeling the unified dark matter and dark energy,
it is often assumed that the parameter $\gamma=1$, that the pressure
of the viscosity fluid is zero and the viscosity term contributes an
effective pressure. There raises some problems here. From the
observational results \cite{w}, the cosmic density nearly equals to
the cosmic pressure. In the viscosity model, the viscosity term
dominates the cosmic pressure, and surpasses the pressure
contributions from other cosmic matter constitutions, which
contradicts the traditional fluid theory. \cite{ni1} \cite{ni2}
propose non-standard interaction mechanism to solve this problem.
Obviously, it is important to build solid foundation for the
research of the viscous cosmology.

Equation of state $w<-1$ lies in the phantom region. It is shown
that cosmology models with such equation of state possess the
so-called the future singularity called the Big Rip \cite{bigrip}.
The larger viscosity model parameter space can help to solve the
cosmic singularity problem and produces different kinds of evolution
mode of the future universe, for more details \cite{sv6}.

The rest of this review is organized as follows: In the next
chapter, general framework of the viscosity model will be reviewed.
In Sec. \textbf{III}, we discuss the modeling of the unified model
with viscosity. In this section, two concrete models are analyzed.
In Sec. \textbf{IV}, data fitting method is introduced briefly.

\section{General framework}
We consider the standard Friedmann-Robertson-Walker metric,
\begin{equation}\label{eq1}
\rd s^2=-\rd t^2+a(t)^2\big(\frac{\rd r^2}{1-kr^2}+r^2\rd
\Omega^2\big).
\end{equation}
For the sake of simplicity, we choose the flat geometry $k=0$, which
is also favored by the update result of the cosmic background
radiation measurement.

The general stress-energy-momentum tensor is
\begin{equation}
T_{\mu\nu}=(\rho+p)U_{\mu}U_{\nu}+pg_{\mu\nu}-\zeta\theta
h_{\mu\nu},
\end{equation}
where $\zeta$ is the bulk viscosity. The expansion factor $\theta$
is defined by $\theta=U^{\mu}_{;\mu}=3\frac{\dot a}{a}$, and the
projection tensor $h_{\mu\nu}\equiv g_{\mu\nu}+U_{\mu}U_{\nu}$. In
the co-moving coordinates, the four velocity $U^{\mu}=(1,0,0,0)$. We
do not specify the concrete form of $\zeta$ in this section.
Generally speaking, $\zeta$ is a quantity evolving with time $t$ or
the scale factor $a(t)$. We will see below that non-trivial and more
complicated $\zeta$ can produce different results especially useful
for the late universe modeling.

From the usual Einstein equation,
\begin{equation}
G_{\mu\nu}=R_{\mu\nu}-\frac{1}{2}Rg_{\mu\nu}=8\pi GT_{\mu\nu},
\end{equation}
we obtain two equations which we call the modified Friedmann
equations:
\begin{subequations}
\begin{eqnarray}
\frac{\dot{a}^2}{a^2} &=& \frac{8\pi G}{3}\rho\label{eq1},\\
\frac{\ddot{a}}{a} &=& -\frac{4\pi
G}{3}(\rho+3\tilde{p})\label{eq2},
\end{eqnarray}
\end{subequations}
where $\tilde{p}$ is an effective pressure,
$\tilde{p}=p-\zeta\theta$.

The covariant conservation equation $T^{0\mu}_{;\mu}=0$ yields
\begin{equation}
\dot\rho+(\rho+\tilde{p})\theta =0.
\end{equation}
The existence of a bulk viscosity contributes a modification to the
pressure $p$, thus we see the Friedmann equation and the covariant
conservation equation are invariant under the transformation
\begin{equation}
p\rightarrow \tilde{p}=p-\zeta\theta,
\end{equation}

The covariant energy conservation equation becomes
\begin{equation}\label{8}
\dot\rho+(\rho-\zeta\theta)\theta=0.
\end{equation}
If define the dimensionless Hubble parameter here
\begin{equation}
h^{2}=\frac{H^2}{H^2_{0}}=\frac{\rho}{\rho_{cr}},
\end{equation}
where $\rho_{cr}=\frac{3H^2_{0}}{8\pi G}$ is the critical density
now. Using the dimensionless Hubble parameter, Eq. (\ref{8}) can be
transformed as
\begin{equation}\label{h}
\frac{1}{H_{0}}\frac{\rd (h^2)}{\rd t}+3h^3=9\lambda h^2,
\end{equation}
where the bulk viscosity is redefined as
$\lambda=\frac{H_{0}\zeta}{\rho_{cr}}$. Through the simple relation
between scale factor $a(t)$ and the redshift $z$
\begin{equation}
\rd t=\frac{1}{aH}\rd a,
\end{equation}
we transform Eq. (\ref{h}) into a differential equation with respect
to the scale factor $a(t)$
\begin{equation}
\frac{\rd H}{\rd a}+\frac{3}{2a}H=\frac{3\zeta}{2a}.
\end{equation}
Solving this equation, we obtain a integral form of $H(a)$
\begin{equation}\label{H}
H(a)=C_{1}a^{-3/2}+\Big[\int\frac{3\zeta}{2a}\mathrm{exp}\big(\int\frac{3}{2a}\rd
a\big)\rd a\Big]\mathrm{exp}\big(-\int\frac{3}{2a}\rd a\big).
\end{equation}
Different forms of viscosity can be used here to make this integral
calculable, numerically or exactly.
\section{Unified single fluid}
\subsection{Redshift-dependent model }
In \cite{hd}, authors assume the bulk viscosity takes the form as an
Hubble parameter dependent function. A redshift-dependent viscosity
is proposed in \cite{sv5}. This bulk viscosity is a combination of a
constant and a scaling relation term
\begin{equation}
9\lambda=\lambda_{0}+\lambda_{1}(1+z)^{n},
\end{equation}
where $n$ is an integer, $\lambda_{0}$ and $\lambda_{1}$ are two
constants, which will be fitted from the observational data sets.

After taking account of this ansatz, the integration is easily to
work out. We get
\begin{equation}
h^{2}(z)=\lambda^2_{2}(1+z)^3+\frac{2}{3}\lambda_{0}\lambda_{2}(1+z)^{1.5}-
\frac{2\lambda_{0}\lambda_{1}}{3(2n-3)}(1+z)^n+\frac{\lambda^2_{1}}{(2n-3)^2}(1+z)^{2n}
-\frac{2\lambda_{1}\lambda_{2}}{2n-3}(1+z)^{n+1.5}+\frac{\lambda^2_{0}}{9}.
\end{equation}
Since we have assumed the spatial flat of the universe, the
consistency condition requires $h(0)=1$. Thus this sets a constraint
on the model parameters as
\begin{equation}
\frac{\lambda_{0}}{3}=1-\lambda_{2}+\frac{\lambda_{1}}{2n-3}
\end{equation}
We remind the readers that we make the single fluid assumption
above, and we do not concretely specify the constitutions of the
cosmic density $\rho$. In this single fluid model, values of model
parameters $\lambda_{0}$, $\lambda_{1}$ and $\lambda_{2}$ will be
fitted, and their meaning are not explained. But when we compare it
with two-fluid model, that the universe is filled with dark matter
and dark energy fluid, more constraints can be added. The solution
consists terms with different scaling relation. The first term has
the form like $C(1+z)^3$, which have the same evolution behavior as
the cold dark matter. Their simplicity leads us to correspond
parameter $\lambda_{2}$ to dark matter ratio $\Omega_{m}$
\begin{equation}
\lambda^{2}_{2}=\Omega_{m}.
\end{equation}
This identity can help us utilize more data to constrain the
viscosity model. The result is also consistent with that obtained
from the standard model($\Lambda$CDM). The shift parameter
$\mathcal{R}$ \cite{sr1} \cite{sr2} and the distance parameter
$\mathcal{A}$ is defined as
\begin{equation}
\mathcal{R}\equiv\sqrt{\Omega_{m}}\int_{0}^{z_{*}}\:\frac{d
z^{'}}{h(z^{'})},
\end{equation}
and
\begin{equation}
\mathcal{A}\equiv\sqrt{\Omega_{m}}\:h(z_{b})^{-\frac{1}{3}}\big(\frac{1}{z_{b}}\int^{z_{b}}_{0}\:\frac{d
z^{'}}{h(z^{'})})^{\frac{2}{3}},
\end{equation}
respectively. Both of them are dependent on dark matter ratio
$\Omega_{m}$, and in the joint statistical analysis they provide
strong constraint on $\Omega_{m}$.

\subsection{Effective equation of state model}
Another viscosity model reviewed here is proposed in \cite{sv3},
where a general form time-dependent viscosity is discussed
\begin{equation}\label{v2}
\zeta=\zeta_{0}+\zeta_{1}\frac{\dot a}{a}+\zeta_{2}\frac{\ddot
a}{\dot a}.
\end{equation}
An interesting feature of this model is its effective equivalence to
the following equation of state
\begin{equation}
p=(\gamma-1)\rho+p_{0}+w_{H}H+w_{H2}H^{2}+w_{dH}\dot H
\end{equation}
where $p_{0}$, $w_{H}$, $w_{H2}$ and $w_{dh}$ are free parameters.
The corresponding between two groups of coefficients are
\begin{subequations}
\begin{eqnarray}
w_{H} &=& -3\zeta_{0},\\
w_{H_{2}} &=& -3(\zeta_{1}+\zeta_{2}),\\
w_{dH} &=& -3\zeta_{2}.
\end{eqnarray}
\end{subequations}
The parameterized bulk viscosity combines terms related to the
``velocity'' $\dot a$ and ``acceleration'' $\ddot a$, which can be
seen to describe the dynamics of the cosmic non-perfect fluid. After
eliminating $p$ and $\rho$, a differential equation about the scale
factor $a(t)$ can be obtained
\begin{equation}
\frac{\ddot
a}{a}=\frac{-(3\gamma-2)/2-(\kappa^{2}/2)w_{H2}+(\kappa^{2})w_{dH}}{1+(\kappa^{2})w_{dH}}\big(\frac{\dot
a}{a}\big)^{2}+\frac{-(\kappa^{2})w_{H}}{1+(\kappa^{2}/2)w_{dH}}\frac{\dot
a}{a}+\frac{-(\kappa^{2}/2)p_{0}}{1+(\kappa^{2}/2)w_{dH}}.
\end{equation}
Another feature of this model is that this differential equation can
be solved exactly, and the evolution function of the scale factor
$a(t)$ is definite. This evolution function is especially convenient
for discussing the cosmic singularity.

With the initial conditions $a(t_{0})=a_{0}$ and
$\theta(t_{0})=\theta_{0}$, when $\tilde{\gamma}\neq 0$, the scale
factor can be obtained as
\begin{eqnarray}
a(t) & = &
a_{0}\big\{\frac{1}{2}\big(1+\tilde{\gamma}\theta_{0}T-\frac{T}{T_{1}}\big)\mathrm{exp}\big[\frac{t-t_{0}}{2}\big(\frac{1}{T}+\frac{1}{T_{1}}\big)\big]+{}
                                                                                                                        \nonumber\\ & &
{}\frac{1}{2}\big(1-\tilde{\gamma}\theta_{0}T+\frac{T}{T_{1}}\big)\mathrm{exp}\big[-\frac{t-t_{0}}{2}\big(\frac{1}{T}-\frac{1}{T_{1}}\big)\big]\big\}^{2/3\tilde{\gamma}}.
\end{eqnarray}
where the parameters are redefined as
\begin{equation}\label{9}
\tilde\gamma=\frac{\gamma+(\kappa^{2}/3)w_{H2}}{1+(\kappa^{2}/2)w_{dH}},
\end{equation}
\begin{equation}
\frac{1}{T_{1}}=\frac{-(\kappa^{2}/2)w_{H})}{1+(\kappa^{2}/2)w_{dH}},
\end{equation}
\begin{equation}
\frac{1}{T^{2}_{2}}=\frac{-(\kappa^{2}/2)p_{0}}{1+(\kappa^{2}/2)w_{dH}},
\end{equation}
\begin{equation}\label{12}
\frac{1}{T^{2}}=\frac{1}{T^{2}_{1}}+\frac{6\tilde\gamma}{T^{2}_{2}}.
\end{equation}
From Friedmann equation, $\rho$ can be written as
\begin{equation}
\rho(t)=\frac{1}{3\kappa^{2}\tilde{\gamma}^{2}}\bigg[\frac{(1+\tilde{\gamma}\theta_{0}T-\frac{T}{T_{1}})(\frac{1}{T}+\frac{1}{T_{1}})\mathrm{exp}(\frac{t-t_{0}}{T})-(1-\tilde{\gamma}\theta_{0}T+\frac{T}{T_{1}})(\frac{1}{T}-\frac{1}{T_{1}})}{(1+\tilde{\gamma}\theta_{0}T-\frac{T}{T_{1}})\mathrm{exp}(\frac{t-t_{0}}{T})+(1-\tilde{\gamma}\theta_{0}T+\frac{T}{T_{1}})}\bigg]^{2}.
\end{equation}
The model parameters leave enough space to produce various evolution
behavior, which can be interpreted in different ways. In this
review, we emphasis its power to unify dark energy and dark matter
with the single fluid assumption. According to the parameters
redefined above and the Friedmann equation, the equation of state
can be converted to
\begin{equation}
p=(\tilde{\gamma}-1)\rho-\frac{2}{\sqrt{3}\kappa
T_{1}}\sqrt{\rho}-\frac{2}{\kappa^{2}T^{2}_{2}},
\end{equation}
The case $\tilde{\gamma}=0$ and $T_{1}\rightarrow\infty$ corresponds
to the $\Lambda$CDM. With the aim to unify dark energy and dark
matter, the case $\tilde{\gamma}=1$ and $T_{2}\rightarrow\infty$ is
especially considered. This case corresponds to a single fluid with
constant viscosity. The relation between $p$ and $\rho$ can be
obtained from the general equation of state above
\begin{equation}
p=-\frac{2}{\sqrt{3}\kappa T_{1}}\sqrt{\rho}.
\end{equation}
Therefore, it is straightforward to eliminate $p$ from the covariant
energy conservation equation, and to work out the solution of
$\rho$. Using Friedmann equation, $H(z)$ can be obtained
\begin{equation}
H(z)=H_{0}[\Omega_{\gamma}(1+z)^{3/2}+(1-\Omega_{\gamma})].
\end{equation}
$\Omega_{\gamma}$ is the only one model parameter. Its value can be
fitted from SNe Ia observational data.
\section{data fitting}
We review the method to fit the model parameters. More details are
illustrated in \cite{fm}. The data sets we use are SNe Ia, BAO and
CMB. The 397 Constitution sample \cite{397} combines the Union
sample \cite{union} and the low redshift ($z<0.08$) sample
\cite{lr}. The co-moving distance $d_{M}$ in FRW coordinate is
\begin{equation}
d_{M}=\int^{z}_{0}\frac{1}{H(z')}\rd z'
\end{equation}
The apparent magnitude which is measured is
\begin{equation}
m\equiv M+5\log_{10}D_{L}(z),
\end{equation}
where the dimensionless luminosity $D_{L}\equiv H_{0}d_{L}(z)$ and
\begin{equation}
d_{L}=(1+z)d_{M}(z).
\end{equation}
where $M$ is the absolute magnitude which is believed to be constant
for all SNe Ia. In the SNe Ia samples, data are given in terms of
the distance modulus $\mu_{obs}\equiv m(z)-M_{obs}(z)$. The $\chi^2$
for this procedure is written as
\begin{equation}
\chi^{2}=\sum_{i=1}^{n}\bigg[\frac{\mu_{obs}(z_{i})-\mu_{th}(z_{i};c_{\alpha})}{\sigma_{obs}(z_{i})}\bigg]^{2}.
\end{equation}
where $\mu_{th}$ means the distance modulus calculated from model
with parameters $c_{\alpha}$ ($\alpha=0,1,2...$). Together with the
shift parameter $\mathcal{R}$ and the distance $\mathcal{A}$, the
total $\chi^{2}_{total}$ for the joint data analysis is
\begin{equation}
\chi^{2}_{total}=\chi^{2}+\left(\frac{\mathcal{R}-\mathcal{R}_{obs}}{\sigma_{\mathcal{R}}}\right)^{2}+\left(\frac{\mathcal{A}-\mathcal{A}_{obs}}{\sigma_{\mathcal{A}}}\right)^{2}.
\end{equation}
For the redshift-dependent model, the relation between distance
modulus and redshift is plotted in FIG.1. The model calculated value
and the Constitution data is compared in the figure.
\begin{figure}
\includegraphics{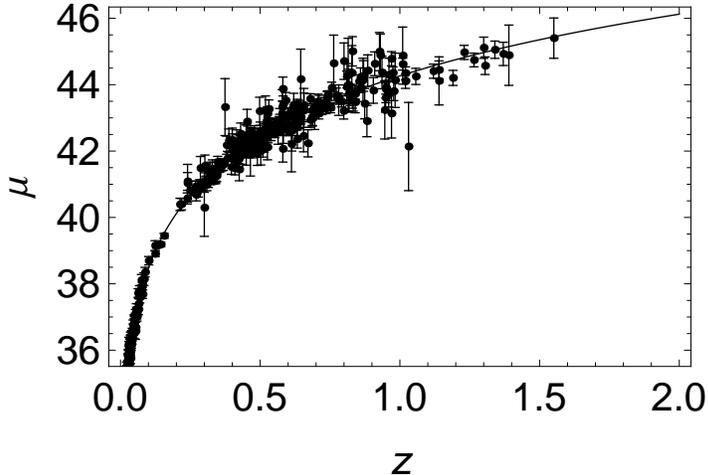}
\caption{Relation between distance modulus and redshift. The solid
line corresponds to the theoretical value calculated from model
concerned. The dots with error bar are the data from the 397
Constitution sample.}
\end{figure}

\section{conclusion}
In this review, we discuss three aspects of the viscosity model,
\begin{itemize}
\item General framework for viscosity modeling. General form of
Hubble parameter is presented. This general form is convenient for
comparing different scale factor(or redshift) dependent viscosity
models.
\item Two kinds of viscosity models are used to model unified
models.
\item Observation constraint is necessary for model building. We can
see the fitting results are consistent with data. It is prospected
that more accurate direct measurements of Hubble constant will
provide a new constraint on cosmological parameters \cite{hp}.
\end{itemize}
Especially we focus on its application on modeling the unified dark
energy and dark matter.

In the cosmic background level, dynamical analysis can be performed.
The statefinder method is useful for discriminating different models
\cite{stf1} \cite{stf2} \cite{stf3} \cite{stf4} \cite{stf5}.
Compared with $\Lambda$CDM model, evolution of the statefinder of
the viscosity model is different and can be discriminated easily,
more details can be found in \cite{stfv1} \cite{sv6}. More plentiful
and accurate data will improve the power of the statefinder method,
which will give enough constraint on the late universe model.

We review the viscosity model which is on the level of zero order.
The perturbation analysis and the large scale structure are
especially useful for the model building. The model predictions need
to be consistent with CMB and LSS data. Some works has investigated
the perturbation aspects of the viscosity model \cite{p1} \cite{v4}.
After corresponding the model parameters, the viscosity model has
the connection with the Chaplygin gas model. Though the Chaplygin
gas model can fit the SNe Ia data well, in the perturbation level it
is found the Chaplygin gas model does not behave in a satisfactory
way. Whether the viscosity models could behave well needs further
investigation.

\section*{Acknowledgements}
This work is supported in part by the National Science Foundation of
China.

\end{document}